\documentclass{article}
\usepackage{ijcai19}

\usepackage{times}
\usepackage{amsmath,bm}
\usepackage{xfrac}
\usepackage{algorithmic}
\usepackage{amsfonts}
\usepackage[inoutnumbered,linesnumbered,ruled,vlined]{algorithm2e}
\usepackage{color}
\usepackage{multirow}
\usepackage{tablefootnote}
\usepackage{url}

\title{
Learning Semantic Annotations for Tabular Data
}
\author{
Jiaoyan Chen$^{1}$ \and Ernesto Jim\'enez-Ruiz$^{2,4}$ \and Ian Horrocks$^{1,2}$ \and Charles Sutton$^{2,3}$
\affiliations
$^1$Department of Computer Science, University of Oxford, UK\\
$^2$The Alan Turing Institute, London, UK\\
$^3$School of Informatics, The University of Edinburgh,  UK\\
$^4$Department of Informatics, University of Oslo, Norway
}

\begin{document}

\maketitle

\begin{abstract}
The usefulness of tabular data such as web tables critically depends on
understanding their semantics.
This study focuses on column type prediction for tables without any meta data.
Unlike traditional lexical matching-based methods,
we propose a deep prediction model that can fully exploit a table's contextual semantics,
including table locality features learned by a Hybrid Neural Network (HNN),
and inter-column semantics features learned by a knowledge base (KB) lookup and query answering algorithm.
It exhibits good performance not only on individual table sets, but also
when transferring from one table set to another.

\end{abstract}

\section{Introduction}
Tabular data such as web tables and legacy databases are a rich and rapidly expanding resource.
They often contain high value data,
but may be hard to use
due to meta data being missing, incomplete or obfuscated. 
Gaining an understanding of their meaning 
is thus of critical importance.
One prominent solution, 
which is often 
referred to 
as semantic table annotation,
is to exploit the semantics of a widely recognized 
knowledge base (KB)
by linking table components, such as columns and cells, to KB components, 
such as classes (categories), entities (elements) and properties (relations).
It can be widely applied in KB population \cite{ritze2016profiling}, search engines \cite{cafarella2008webtables,cafarella2018ten}, automatic data analysis \cite{thirumuruganathan2018data,chu2015katara} and so on.

Semantic table annotation has been extensively studied, especially for web tables \cite{cafarella2018ten}.
Traditional methods are mostly based on lexical matching by name, 
with annotation modeled as tasks such as matching cells to entities, columns to classes, inter-column relations to properties and so on \cite{limaye2010annotating}.
Other methods, including probabilistic graphical models \cite{bhagavatula2015tabel} and iterative algorithms \cite{ritze2015matching}, have been developed to explore the correlation between different matching tasks for disambiguation.
However, most of them rely on table metadata such as column names to jointly model multiple matching tasks,
while lexical matching itself fails to capture the contextual semantics of a name.

Recently some studies have explored the use of deep learning in semantic table annotation.
For example
\cite{luo2018cross} learns 
cell contextual
features to predict its corresponding KB entity
(cf. Section \ref{sec:related_work}).
These works illustrate the benefit of deep learning in modeling contextual semantics of tables,
but they still have limitations:
\textit{(i)} some tasks, such as column type annotation, have not been fully investigated;
\textit{(ii)} some contextual semantics, such as inter-column relations, have not been fully explored; and 
\textit{(iii)} the transferability (generalization) of the learned model has not been evaluated.

In this study, we focus on semantic type (i.e., class) prediction for columns that are composed of phrases (i.e., entity mentions).
For example, a column composed of ``Google'', ``Amazon'' and ``Apple Inc.'' 
can be annotated by the class \textit{Company}.
To this end, we first develop a Hybrid Neural Network (HNN)
to model the contextual semantics of a column.
It embeds the phrase within a cell
with a bidirectional Recurrent Neural Network and an attention layer (Att-BiRNN),
and learns \textit{(i)} column features (i.e., intra-column cell correlation)
and \textit{(ii)} row features (i.e., intra-row cell correlation)
with a Convolutional Neural Network (CNN).

The arbitrary relative position of columns makes it difficult for the neural network to learn general row features.
Thus we extend the row features with property features, which indicate potential relations between 
columns and provide discriminative predictive information.
For example, given a column composed of ``Animal Farm'', ``The Goldfather'' and ``Brokeback Mountain'',
together with a column of person names, 
the potential relation \textit{director} indicates the first column is more likely to be of type \textit{Film}, 
while the relation \textit{author} suggests \textit{Book} as probable type. 
To extract such property features, a novel KB lookup and reasoning algorithm was developed.

In summary, this study contributes a new column type prediction method combing HNN for feature learning and KB lookup and reasoning for feature extraction.
We evaluate our technique using the DBpedia KB and three table sets:
T2Dv2 from the general Web, 
Limaye and Efthymiou from 
the Wikipedia encyclopedia.
As well as testing single table sets, 
the evaluation specially considers the generalization (transferability) of the prediction model from one table set to another.
The evaluation suggests that our method is effective and that its overall accuracy is higher than the state-of-the-art in most cases.


\section{Methodology}

\subsection{Problem Statement}\label{sec:problem_statement}

We assume a table is composed of cells organized by columns and rows, 
without any metadata like column names.
The input is a table with a \textit{target column} whose type is to be predicted.
The column includes ordered cells, 
each of which is a sequence of words (text phrase), known as an \textit{entity mention}.
A column composed of entity mentions is also known as an \textit{entity column}.
Other columns in the input table are called \textit{surrounding columns}.
We assume a fixed set of candidate classes that are disjoint with each other are given, denoted as $\left\{C_1, ..., C_K \right\}$.
The problem is assigning a real value score to each candidate class so that the correct class (type) of the target column has the highest score.

The input of our method is modeled as a fixed structure table called a \textit{micro table}, denoted as $S$.
It has one target column with a fixed number of cells, denoted as $\mathcal{L} = (\mathcal{L}_{1}, ..., \mathcal{L}_{m} )$,
and a fixed number of surrounding columns, denoted as $\bm{L} = ( L_{1}, ..., L_{l} )$.
The first cell of the target column $\mathcal{L}_1$ is known as the micro table's \textit{main cell}.

In training,
we assume $n$ labeled micro tables (samples) are 
%
extracted from labeled entity columns by \textit{(i)} sliding a window from the first row to the last with the step of one cell and \textit{(ii)} selecting surrounding columns from the left to the right.
A function (model) $\mathcal{F}: S \rightarrow y$ is learned, 
where $y \in R^K$ represents the output vector. 
In predicting the type of a target column with size $M$, 
micro tables are first extracted 
and predicted by the trained model $\mathcal{F}$. 
Their output vectors are then averaged as the final score vector to the target column: 
$\bar{y} = \frac{1}{M-m+1} \sum_{i=1}^{M-m+1} \mathcal{F}(S_i)$.
The remainder of this section presents our model $\mathcal{F}$, 
while some of its training details are presented in Section \ref{sec:experiment_settings}.

\subsection{HNN Architecture}
\begin{figure}[h]
\centering
\includegraphics[scale=0.364]{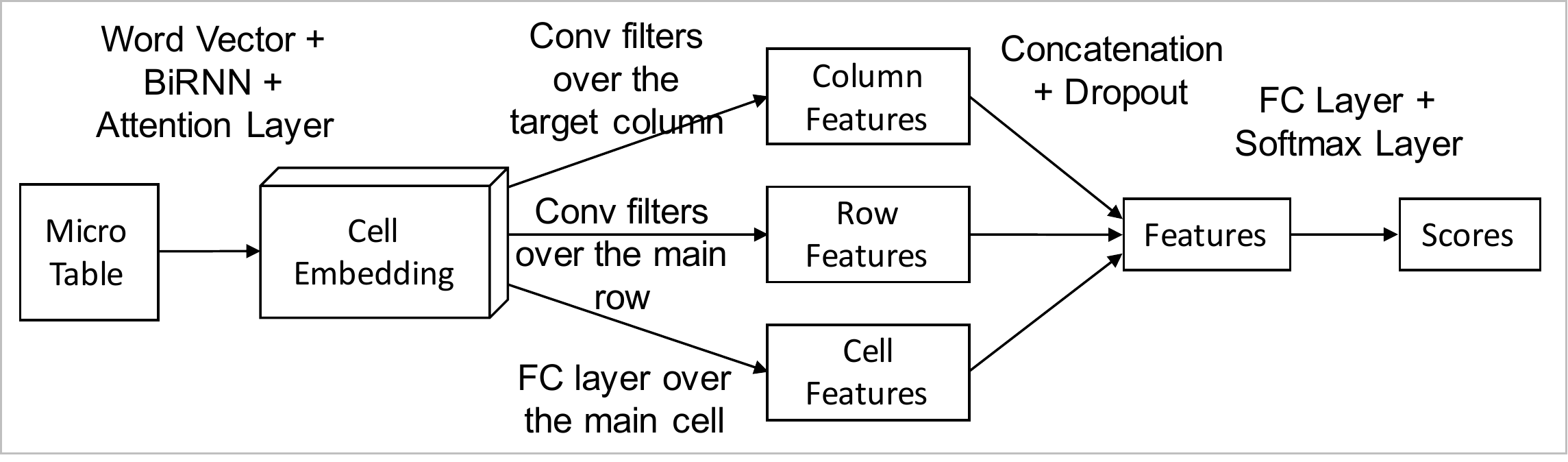}
\vspace{-0.35cm}
\caption{A brief view of the HNN architecture.
}
\label{fig:hnn}
\end{figure}

Our HNN mainly includes an attentive BiRNN for cell embedding,
and a customized convolutional (Conv) layer for table locality feature learning, 
as shown in Figure \ref{fig:hnn}.
\vspace{-0.2cm}
\vspace{-0.2cm}

\subsubsection{Cell Embedding}
We use an RNN with Gated Recurrent Unit (GRU) \cite{bhagavatula2015tabel} to embed the word sequence of each cell ($x_t, t \in \left[1,T \right]$).
It uses 
a reset gate $r_t$ to control the contribution of past state (word), 
and an update gate $z_t$ to balance the contributions of past information and new information.
The hidden state at position $t$ is computed as
\begin{equation}\label{eq:rnn1}
h_t = (1-z_t) \odot h_{t-1} + z_t \odot \tilde{h}_t,
\end{equation}
where $\odot$ denotes the Hadamard product, 
$h_{t-1}$ represents the past state,
$\tilde{h}_t$ is a state computed with new sequence information.
$\tilde{h}_t$, $z_t$ and $r_t$ are updated as
\vspace{-0.1cm}
\begin{equation}\label{eq:rnn2}
\begin{cases}
\tilde{h}_t = \text{tanh}(W_h x_t + r_t \odot (U_h h_{t-1}) + b_h), \\
z_t = \sigma(W_z x_t + U_z h_{t-1} + b_z), \\
r_t = \sigma(W_r x_t + U_r h_{t-1} + b_r).
\end{cases}
\vspace{-0.1cm}
\end{equation}

Assume the cell phrase length is fixed to $T$ by cropping and padding,
and each cell phrase is represented as $(v_1,...,v_T)$ where $v_t$ denotes the vector of the word at position $t$.
In BiRNN, 
both forward hidden states ($\overrightarrow{h_t} = \overrightarrow{\text{GRU}}(v_t)$, $t \in \left[1,T \right]$) and
backward hidden states ($\overleftarrow{h_t} = \overleftarrow{\text{GRU}}(v_t)$, $t \in \left[T,1 \right]$) are calculated.
The embedding of the word at position $t$, denoted as $e_t$, 
is the concatenation of $\overrightarrow{h_t}$ and $\overleftarrow{h_t}$.

The embedding of a cell phrase is composed of the BiRNN embeddings of its words.
Inspired by \cite{yang2016hierarchical}, 
we assume different words are differently informative towards a prediction task,
and an attention layer is thus stacked.
Given a phrase with BiRNN word embedding  ($e_t, t \in \left[1,T \right]$), 
the attention layer output is
$a = \sum_{t} \alpha_t e_t$, where $\alpha_t$ is the normalized weight of the word at position $t$ and is calculated as 
\vspace{-0.1cm}
\begin{equation}\label{eq:att}
\begin{cases}
\alpha_t = \frac{\text{exp}(u^T_t u_w)}{\sum_t \text{exp} (u^T_t u_w)} \\
u_t = \text{tanh} (W_w e_t + b_w) 
\end{cases}
\vspace{-0.02cm}
\end{equation}
The dimension of cell embedding $a$ is denoted as $d_0$;
$u_w$ represents the informative degree of all the words in training.

Given a micro table, the cells of the target column and the cells of its surrounding entity columns are embedded by the above Att-BiRNN;
the cells of the surrounding real value columns are transformed into a vector of dimension $d_0$ by zero padding;
the cells of the surrounding date columns are first parsed with integers of year, month and day,
and then transformed into a vector of dimension $d_0$ by concatenating the integers and zero padding.

One column is embedded into a matrix of size $m \times d_0$ by stacking vectors of its cells.
For convenience, we also use the annotation of a column 
to denote its embedded matrix (i.e., $\mathcal{L}$ for the target column, $L_i$ for a surrounding column).
One micro table is embedded into a tensor of size $m \times (l+1) \times d_0$ by stacking matrices of its columns, denoted as $\left[\mathcal{L}, L_1, ..., L_l \right]$.

\subsubsection{Column Features and Row Features}
One Conv layer is stacked after Att-BiRNN,
including
\textit{(i)}~Conv filters over the target column for column feature learning, denoted as $c_1$, 
and \textit{(ii)} Conv filters over the row of the main cell for row feature learning, denoted as $c_2$. 

Each filter over the column $W^{c_1}_i$ has the size of $k_1  \times d $, where $k_1 \in \Theta_1$, $\Theta_1 \subseteq \left\{ 2,...,m \right\}$.
Given the matrix of the target column $\mathcal{L}$, 
the filter computes the column features as
\begin{equation}
f^{c_1,k_1}_i = g(W^{c_1}_i \otimes \mathcal{L} + b^{c_1}),
\end{equation}
where $\otimes$ denotes the Conv operation, 
$g$ denotes an activation function like ReLu
and $b^{c_1}$ denotes the biases.

Each filter over the row 
$W^{c_2}_j$ has the size of $1 \times k_2 \times d$, 
where $k_2 \in \Theta_2$, $\Theta_2 \subseteq \left\{ 2,...,l+1 \right\}$.
Given the tensor of a micro table,
the filter computes the row features as 
\begin{equation}
    f^{c_2,k_2}_j = g(W^{c_2}_j \otimes \left[ \mathcal{L}_1, L_{1,1},...,L_{l,1} \right] + b^{c_2}),
\end{equation}
where $L_{i,1}$ denotes the first cell of surrounding column $L_i$, $b^{c_2}$ denotes the biases.
It models the correlation between the target column and its surrounding columns.

Inspired by some successful CNN architectures with one Conv layer (e.g., \cite{kim2014convolutional} for text classification),
a max pooling layer is stacked after the Conv layer to extract salient signals and regularize the network.
Thus the column filter $k_1 \times d$ finally computes the output as
\begin{equation}\label{eq:max-pooling}
    f^{c_1, k_1} = [\text{max}(f_1^{c_1, k_1}),\text{max}(f_2^{c_1, k_1}), ..., \text{max}(f_{\kappa_1}^{c_1, k_1})] \\
\end{equation}
where $\text{max}(\cdot)$ denotes a vector's maximum value, 
$\kappa_1$ denotes the number of features to be learned for each filter.
For the row filter $1 \times k_2 \times d$, with the number of features $\kappa_2$,
the output,
denoted as $f^{c_2, k_2}$ 
is calculated in the same way as \eqref{eq:max-pooling}.

The max pooling layer concatenates $f^{c_1,k_1}$, $k_1 \in \Theta_1$ and $f^{c_2,k_2}$, $k_2 \in \Theta_2$ as the output, denoted as $f^{c_1,c_2}$.
$\Theta_1$, $\Theta_2$, $\kappa_1$ and $\kappa_2$ are hyper parameters about the HNN architecture.

A fully connected (FC) layer is then stacked for modeling the nonlinear relationship.
It calculates the output as
\vspace{-0.05cm}
\begin{equation}\label{eq:f_hnn}
f^{hnn} = f^{c_1,c_2} \cdot W^{fc} + b^{fc},    
\end{equation}
where $\cdot$ denotes matrix multiplication,  
$W^{fc}$ and $b^{fc}$ denote weights and biases of the FC layer.
Finally, a softmax layer is stacked to calculate the output score for each class:
\vspace{-0.15cm}
\begin{equation}\label{eq:y_hnn}
y_i^{hnn} = \sfrac{\text{exp}(f_i^{hnn})}{\sum_{j=1}^K \text{exp}(f_j^{hnn})},
\vspace{-0.25cm}
\end{equation}
where $i=1,2,...,K$.

\subsection{Property Features}
Property features are used to represent the potential relations between the target column and its surrounding columns.
We first introduce some KB background 
and then present how property features are extracted and incorporated.

\subsubsection{RDF-based Knowledge Base}
The KB in this study follows Semantic Web standards including RDF (Resource Description Framework), RDF Schema, OWL (Web Ontology Language) and SPARQL \cite{domingue2011handbook}.
One KB is composed of a TBox (terminology) and an ABox (assertions).
The TBox, often using RDF Schema, contains constructors for
the definition of class,
class relations (e.g., \textit{rdfs:subClassOf} for the descendent relation),
property,
property domain and range, etc.
It can also use more expressive languages such as OWL 
with 
more powerful constructs such as relation composition.
The ABox contains entities, each of which is represented by an URI (Uniform Resource Identifier),
and RDF triples 
$\left\langle s, p, o \right\rangle$,
where $s$ represents a subject (an entity), 
$p$ represents a predicate (a property) 
and $o$ represents an object
(either an entity or a data value like date and number).
An entity can belong to
one or more classes, which is defined by the property \textit{rdf:type}. 

Such a KB is often called an RDF-based KB. 
It can be accessed by SPARQL queries.
Two examples used in our method are ($\text{\textit{Q}}_\text{\textit{1}}$) getting entities of a given class according to \textit{rdf:type},
and ($\text{\textit{Q}}_\text{\textit{2}}$) getting triples whose subject entity is given.
SPARQL supports semantic reasoning for accessing implicit knowledge \cite{glimm2012sparql};
for example, inferring $\left\langle e \text{ \textit{rdf:type} } c_2 \right\rangle$,
given $\left\langle e \text{ \textit{rdf:type} } c_1 \right\rangle$ and $\left\langle c_1 \text{ \textit{rdfs:subClassOf} } c_2 \right\rangle$.
A KB can also be accessed via fuzzy matching,
with a lexical index on entity labels (phrases defined by \textit{rdfs:label}) and sometimes entity anchor text (short descriptions).
This is often referred to as KB lookup.
Successful systems include Spotlight for DBpedia \cite{mendes2011dbpedia} and  OpenRefine for Wikidata \cite{ham2013openrefine}.

\subsubsection{Candidate Properties}
Given a class $c$ defined by a KB,
we denote entities that belong to it as $E(c)$.
It means the triple $\left\langle e \text{ \textit{rdf:type} } c\right\rangle$ is true 
for any entity $e$ in $E(c)$.
Given a property $p$ defined by a KB,
an entity is defined as a \textit{subject entity} of $p$, denoted as $e_p$, 
if there exists at least one object $o$ such that the triple $\left\langle e_p, p, o \right\rangle$ is entailed by the KB.
We denote all the subject entities of the property $p$ as $E(p)$.
A property is defined as a \textit{frequent property} of class $c$, denoted as $p_c$,
if $\sfrac{\left\vert E(c) \cap E(p_c) \right\vert}{\left\vert E(c) \right\vert} \ge \sigma$,
where $\sigma \in [0,1]$ is a threshold and $\left\vert \cdot \right\vert$ denotes the cardinality of a set.
``Frequent'' means at least a specified proportion of the entities of a class are associated to that specific property.

A candidate property represents a potential relationship between two columns.
To get candidate properties, 
we first extract the frequent properties of each candidate (training) class $C_i \in \mathcal{C}$, denoted as $\bm{p}_{i}$,
and then merge these frequent properties: 
$\bm{P} = \cup_{i=1}^{K} \bm{p}_{i}$.
The size of $\bm{P}$ is denoted as $d_1$.
The above calculation requires $K$ SPARQL queries of type $\text{\textit{Q}}_\text{\textit{1}}$ 
and $\left\vert \cup_{i=1}^K E(C_i) \right\vert$ SPARQL queries of type $\text{\textit{Q}}_\text{\textit{2}}$.


\subsubsection{Property Vector (P2Vec)}
Property features of one micro table are represented by a P2Vec denoted as $v$.
Each slot of $v$ represents the degree of existence of one candidate property, 
and thus the dimension of $v$ is $d_1$.
The calculation of P2Vec is shown in Algorithm \ref{alg:prop_vec}.
Given a micro table, it first retrieves KB entities that match the main cell (Line \ref{alg_line:lookup}).
As lookup by lexical matching is ambiguous,
\textit{entity\_lookup} is set to return more than one entity (at most $\mathbb{N}$) to avoid missing the right entity. 
For each matched entity, it first retrieves its property annotations, namely the triples whose subject is this entity, using a SPARQL query of type $\textit{\textit{Q}}_\text{\textit{2}}$ (Line \ref{alg_line:foreach1} to \ref{alg_line:sparql}),
and then matches each triple's object with the first cell of each surrounding column (Line \ref{alg_line:foreach2} to \ref{alg_line:match}).

In matching, the function \textit{cell\_object\_match} first classifies the object $o$ into types of entity, date, text and number, 
and then returns true or false with the following processing.
An entity is transformed to a phrase with its English label defined by \textit{rdfs:label},
while 
a date is transformed to an integer that represents the year.
In comparing two texts, it returns true if their string-edit distance (e.g., Jaro Distance \cite{cohen2003comparison}) exceeds the threshold $\alpha$ and false otherwise,
while in comparing two numbers, it returns true if they are equal and false otherwise. 
Note that we do not
return a 
matching degree score but true or false, 
so as to leave salient predictive information about inter-column relations with less noise.

Algorithm \ref{alg:prop_vec} needs once entity lookup, at most $\mathbb{N}$ SPARQL queries of type \textit{Q2}, 
and $\mathbb{N} \times d_1 \times l$ matchings with function \textit{cell\_object\_match}.



\begin{algorithm}[h!]
\small
\KwIn{
\textit{(i)} A micro table $(\mathcal{L},\bm{L})$,
\textit{(ii)} candidate properties $\bm{P}$ with the size of $d_1$,
\textit{(iii)} a maximum number of matched entities $\mathbb{N}$,
\textit{(iv)} a text matching threshold $\alpha$,
} 
\KwResult{
$v$: a property vector of the micro table
}
\Begin{
$v := \text{\textit{zeros}}(d_1)$; \emph{$\%$ Init. of the property vector}\\
$E := \text{\textit{entity\_lookup}} (\mathcal{L}_1, \alpha)$; \emph{$\%$ Entity lookup by main cell} \label{alg_line:lookup}\\
\ForEach{entity $e \in E$ \label{alg_line:foreach1}}{
$T := \text{\textit{query}(e)}$;\emph{$\%$ Get triples whose subject is $e$} \label{alg_line:sparql}\\
\ForEach{triple $(s, p, o) \in T$ with $p \in \bm{P}$ \label{alg_line:foreach2}}{
\ForEach{surrounding column $L_i \in \bm{L}$ \label{alg_line:foreach3}}{
\If{$\text{\textit{cell\_object\_match}}(L_{i,1}, o, \alpha)$\label{alg_line:match}}{
$j := \text{\textit{index}}(p, \bm{P})$; \\
$v[j] := 1$; \emph{$\%$ Set the slot of the property} \\
}
}
}
}
$v := \sfrac{v}{\left\| v \right\|}$; \emph{$\%$ Normalization}
}
\Return{$v$}
\caption{ 
\label{alg:prop_vec} 
\small{\tt{P2VecExtract} $\langle (\mathcal{L},\bm{L}), \bm{P}, \mathbb{N}, \alpha \rangle$}
}

\end{algorithm}

\subsection{Ensemble}
P2Vec is integrated with the HNN by two ensemble approaches.
Ensemble I first trains a basic multi-class classifier e.g., Multiple Layer Perception (MLP) and predicts the score:
\begin{equation}\label{eq:ensemble1_1}
y^{p2vec} \xleftarrow{\text{classifier e.g., MLP}} \left[ \mathcal{L}_1, v \right],
\end{equation}
where the average word vector of the main cell $\mathcal{L}_1$ is concatenated with the P2Vec $v$ as the input.
It then calculates the average of the above score and the score by the HNN \eqref{eq:y_hnn}: 
\vspace{-0.1cm}
\begin{equation}\label{eq:ensemble1_2}
y = \sfrac{(y^{hnn} + y^{p2vec})}{2}.
\end{equation}

Ensemble II trains a multiple-class classifier with the concatenation of the P2Vec $v$ and the FC layer output of the HNN~\eqref{eq:f_hnn},
and predicts the score: 
\begin{equation}\label{eq:ensemble2}
y \xleftarrow{\text{classifier e.g., MLP}} \left[f^{hnn}, v \right].
\end{equation}
In decision making, the class with the highest score is adopted as the column type.

\section{Evaluation}\label{sec:experiment_settings}
In the evaluation\footnote{Codes: https://github.com/alan-turing-institute/SemAIDA} conducted in this paper we rely on DBpedia and three web table sets: T2Dv2\footnote{http://webdatacommons.org/webtables/goldstandardV2.html} from the general Web, Limaye \cite{limaye2010annotating} and Efthymiou \cite{efthymiou2017matching} from the Wikipedia encyclopedia.
We annotate \textit{(i)} $411$ entity columns of T2Dv2 with $37$ concrete and disjoint classes defined by the DBpedia ontology,
\textit{(ii)} $114$ entity columns of Limaye with $8$ out of the above $37$ classes,
and \textit{(iii)} $620$ entity columns of Efthymiou with $31$ out of the above $37$ classes.
T2Dv2 is randomly split into T2D-Tr ($70\%$) and T2D-Te ($30\%$). 
All the results except for Table \ref{res:baseline} are based on the following setting: 
T2D-Tr is used for training, 
while T2D-Te, Limaye and Efthymiou
are used as three testing sets.
We report accuracy, i.e., the ratio of correctly labeled columns.



The reported results are based on the following hyper parameter setting.
Regarding the micro table, 
the number of rows $m$ is set to $5$,
the number of surrounding columns $l$ is set to $4$,
and zero-padding is used for tables that do not have enough columns or rows.
In training, 
negative samples are constructed by labeling the entity column with each wrong class;
a word2vec model \cite{mikolov2013distributed} trained by the latest dump of Wikipedia articles is adopted.
HNN is trained by Adam \cite{kingma2014adam} with the loss function of softmax cross entropy.
The hidden size and the attention layer size of RNN are set to $150$ and $50$,
the column Conv filter set $\Theta_1$ and the row Conv filter set $\Theta_2$ are set to $\left\{2,3,4\right\}$ and $\left\{2,3\right\}$,
the feature number per filter ($\kappa_1$ and $\kappa_2$) is set to~$32$.
In computing P2Vec, the
DBpedia lookup service\footnote{https://github.com/dbpedia/lookup} and SPARQL endpoint\footnote{http://dbpedia.org/sparql} are used,
while the hyper parameters $\sigma$, $\mathbb{N}$ and $\alpha$ are set to $0.005$, $5$ and $0.85$ respectively.

In evaluation, we adopt as baselines two typical multi-class classifiers 
-- Logistic Regression (LR) and Multiple Layer Perception (MLP),
variants of our HNN (including ColNet \cite{chen2019colnet}), 
and two lexical matching based column type annotation methods --
DBpedia lookup service plus majority voting by matched entities \cite{zwicklbauer2013towards} (Lookup-vote) and T2K Match \cite{ritze2015matching}.
LR and MLP are also used as the classifier for ensemble.
In the following, we first consider the effectiveness of HNN and P2Vec
and then evaluate the overall result,
with the transferability between table sets analyzed.

\subsection{Hybrid Neural Network}
In Table \ref{res:hnn}, we can see that the HNN variants with both Att-BiRNN and CNN achieve the highest accuracy on all three testing sets.
In the following, we separately analyze the impact of Att-BiRNN and CNN.

\vspace{0.1cm}
\noindent \textbf{Att-BiRNN.}
In comparison with word vector averaging, embedding the cell phrase by Att-BiRNN improves the model's accuracy.
In Table \ref{res:hnn}, Att-BiRNN outperforms word2vec-avg + FC-Softmax by $3.2\%$, $6.4\%$ and $11.3\%$ on T2D-Te, Limaye and Efthymiou respectively.
When a CNN is stacked,
embedding by Att-BiRNN is still beneficial. 
For instance, Att-BiRNN + $\text{CNN} ^ {\text{cr}}$
outperforms word2vec-avg + $\text{CNN} ^ {\text{cr}}$ by $2.5\%$, $9.1\%$ and $9.4\%$ on the three testing sets.

\begin{table}[h!]
\scriptsize{
\centering
\begin{tabular}[t]{c|c|c|c}
\hline
Methods & T2D-Te & Limaye & Efthymiou    \\\hline 
word2vec-avg + FC-Softmax & $0.925$  & $0.561$ & $0.582$ \\\hline
word2vec-avg + $\text{CNN}^{\text{c}}$ & $\bm{0.947}$  & $0.597$ & $\bm{0.619}$  \\
word2vec-avg + $\text{CNN}^{\text{r}}$ & $0.872$  & $\bm{0.675}$ & $0.460$ \\
word2vec-avg + $\text{CNN}^{\text{cr}}$ & $0.902$ & $0.667$ & $0.531$  \\\hline\hline
Att-BiRNN &$0.955$  & $0.597$ & $0.648$ \\\hline
Att-BiRNN + $\text{CNN}^{\text{c}}$ &$\bm{0.962}$  & $0.632$ & $\bm{0.655}$  \\
Att-BiRNN + $\text{CNN}^{\text{r}}$ &$0.880$  & $0.684$ & $0.529$  \\
Att-BiRNN + $\text{CNN}^{\text{cr}}$ & $0.925$ & $\bm{0.728}$  & $0.581$ \\\hline
\end{tabular}
\vspace{-0.2cm}
\caption{\footnotesize
\label{res:hnn}
Accuracy of HNN variants. 
word2vec-avg represents averaging the word2vec of words of each cell phrase.
FC-Softmax denotes a classifier by a FC layer and a Softmax layer.
The superscripts c and r of CNN denote Conv filters over the column and row.
}
}
\end{table}
\vspace{-0.2cm}

\noindent \textbf{Column Features.}
Conv filters over the target column learn column features.
According to Table \ref{res:hnn}, they are effective in improving the accuracy.  
For example, word2vec-avg + $\text{CNN}^{\text{c}}$ outperforms word2vec-avg + FC-Softmax  
by $2.4\%$, $6.4\%$ and $6.4\%$ on T2D-Te, Limaye and Efthymiou respectively.
When the embedding by Att-BiRNN is used, 
they are still beneficial. 
The corresponding improvement of Att-BiRNN + $\text{CNN}^{\text{c}}$ over Att-BiRNN + FC-Softmax is $0.7\%$, $5.9\%$ and $1.1\%$.
The limited improvement on T2D-Te is due to the high base accuracy (T2D-Te comes from the same table set as the training data).

\begin{figure}[h]
\centering
\includegraphics[scale=0.365]{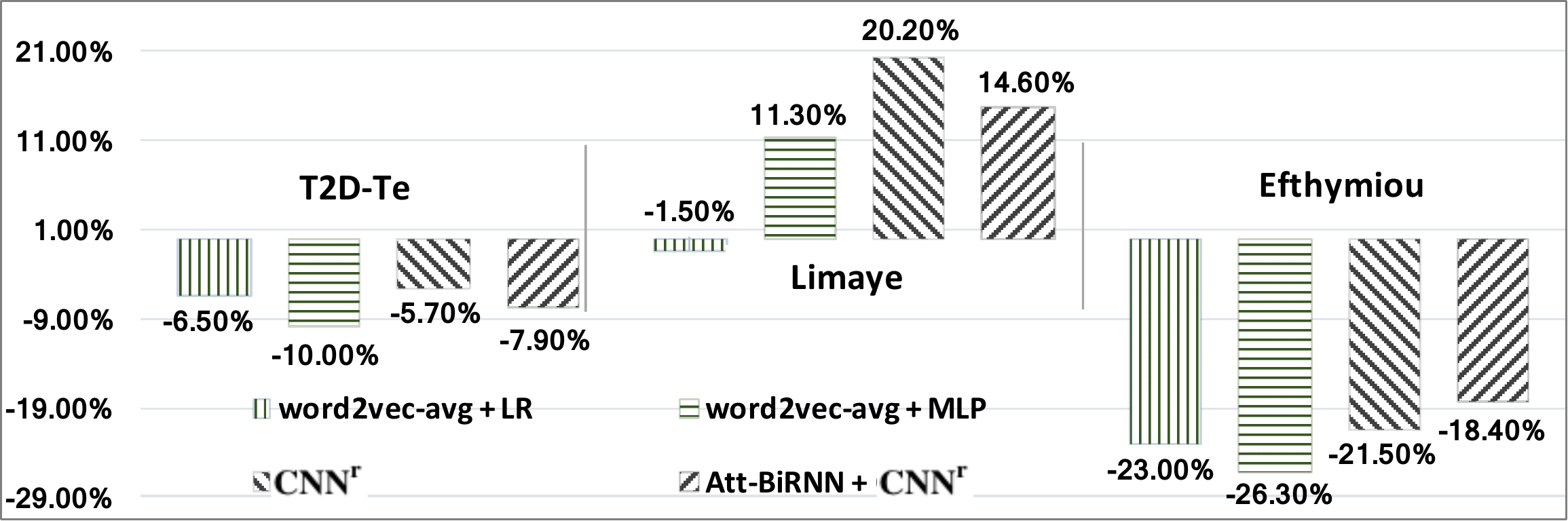}
\vspace{-0.7cm}
\caption{\footnotesize
Accuracy improvement using surrounding columns.
Cells of surrounding column are appended to the main cell through vector concatenation (word2vec-avg + LR and word2vec-avg + MLP) and row feature learning ($\text{CNN}^{\text{r}}$ and Att-BiRNN + $\text{CNN}^{\text{r}}$).
}
\label{res:row}
\end{figure}

\vspace{0.1cm}
\noindent \textbf{Row Features.}
Conv filters over the row of the main cell learn row features.
Unlike column features, the impact of row features varies from data to data, as seen in Table \ref{res:hnn}.
For example, with Att-BiRNN,
adding $\text{CNN}^{\text{r}}$ improves the accuracy by $14.6\%$ on Limaye 
but reduces the accuracy by $7.9\%$ and $18.4\%$ on T2D-Te and Efthymiou respectively.
One potential reason is that the noise from surrounding columns overwhelms the learned discriminative patterns 
due to factors like varying relative position between a target column and a surrounding column (e.g., a column of book names vs a column of writer names) from table to table.
%
This explanation is supported by the results in Figure \ref{res:row} using the basic classifiers LR and MLP.
As with adding row feature via CNN, 
concatenating the average word2vec of cells of surrounding columns increases the accuracy on Limaye but reduces the accuracy on T2D-Te and Efthymiou.

Although row features do not always improve the accuracy, they are still beneficial in comparison with directly concatenating the average word2vec of cells of surrounding columns, leading to higher improvement on Limaye and lower decreasement on T2D-Te and Efthymiou, as seen in Figure \ref{res:row}.


\vspace{-0.15cm}
\vspace{-0.1cm}

\subsection{Property Vector}
\begin{figure}[h]
\centering
\includegraphics[scale=0.34]{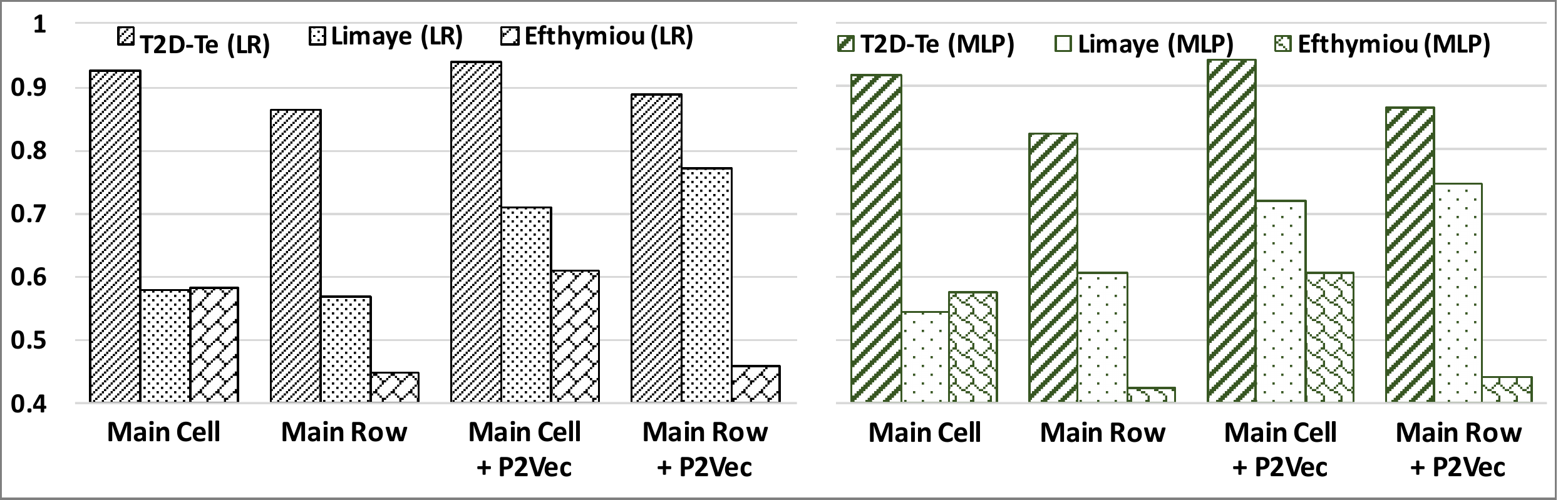}
\vspace{-0.6cm}
\caption{\footnotesize
Accuracy with and without P2Vec concatenation.
Average word2vec is used for cell embedding.
}
\label{res:p2vec1}
\end{figure}

The results in Figure \ref{res:p2vec1} illustrate the effectiveness of P2Vec in column type prediction.
%
On the one hand, appending P2Vec to the main cell (i.e., Main Cell + P2Vec) significantly improves accuracy; e.g., the improvement of MLP is $2.3\%$, $32.2\%$ and $5.2\%$ on T2D-Te, Limaye and Efthymiou respectively.
This is much higher than directly concatenating average word vectors of cells of surrounding columns in the row (i.e., Main Row).
The latter actually negatively impacts performance on T2D-Te and Efthymiou,
which is consistent with the impact of row features learned by Conv filters in HNN.
On the other hand, we find that feeding LR and MLP with P2Vec concatenation even outperforms the HNN that learns row feature.
For example, Main Cell + P2Vec with LR in Figure \ref{res:p2vec1} outperforms Att-BiRNN + $\text{CNN}^{\text{r}}$ in Table \ref{res:hnn} by $6.4\%$, $3.9\%$ and $15.5\%$ respectively on T2D-Te, Limaye and Efthymiou.

\vspace{-0.18cm}
\vspace{-0.17cm}

\begin{figure}[h]
\centering
\includegraphics[scale=0.4]{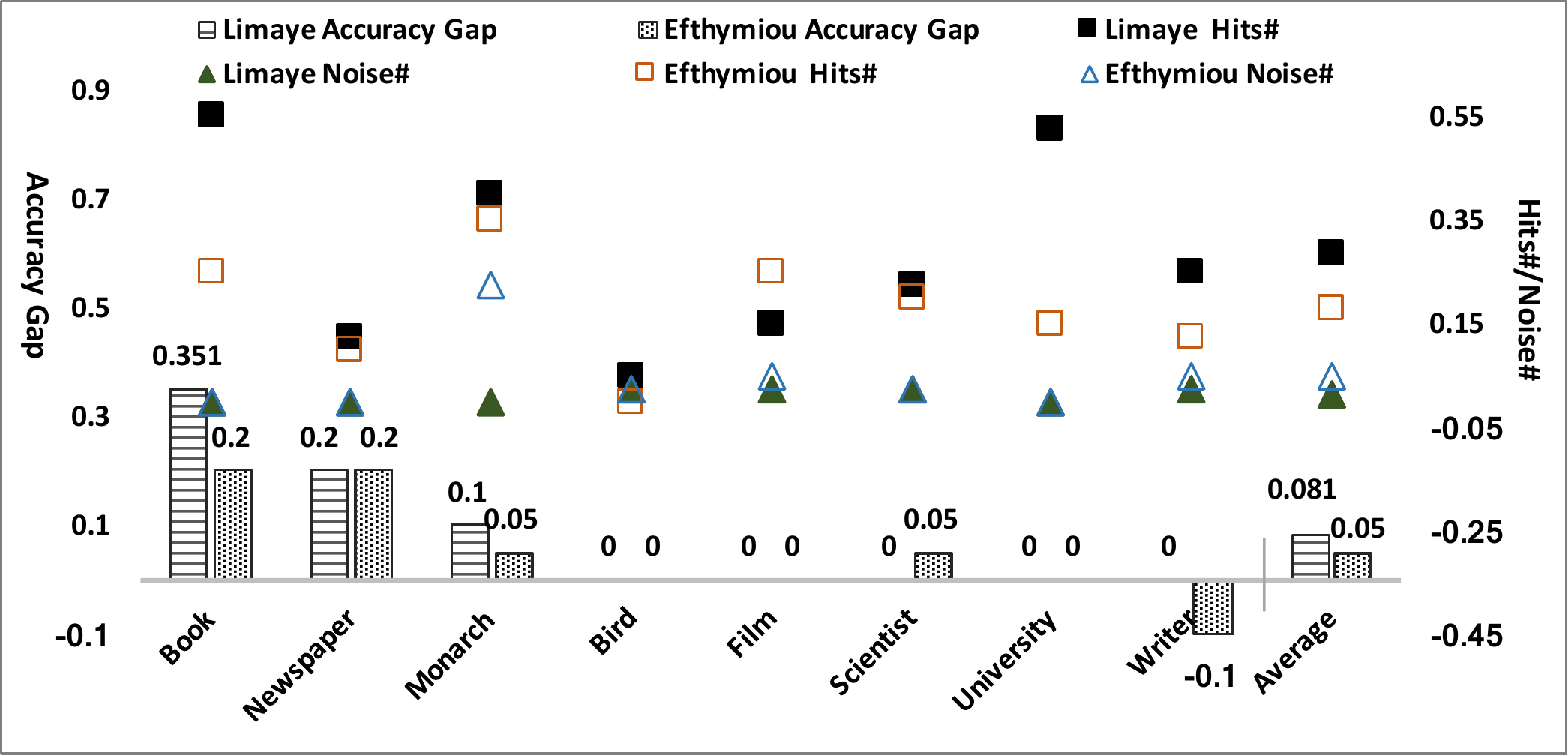}
\vspace{-0.68cm}
\caption{\footnotesize
Average number of correctly and incorrectly matched properties per row, i.e., correct and incorrect none zero elements per P2Vec (Hits\# and Noise\#),
and the accuracy improvement (gap) of LR by appending P2Vec to the main cell,
on $8$ classes.
}
\label{res:p2vec2}
\end{figure}

In Figure \ref{res:p2vec2} we analyze the distribution of non-zero elements of P2Vec 
and its impact on performance improvement by P2Vec.
P2Vec shows significant performance improvement on ``Book", ``Newspaper" and ``Monarch",
and at the same time has significant Hits\# and zero Noise\# (except for ``Monarch" of Efthymiou). 
This indicates the positive impact of the correctly matched properties.
Meanwhile we find there are no or limited improvements on the other $5$ classes although most of them also have significant Hits\#.
This is due to
\textit{(i)} the high base accuracy without P2Vec 
(e.g., close to $1$ for ``Bird" and ``University" of Limaye and Efthymiou), 
and \textit{(ii)} the negative impact of Noise\# 
(e.g., ``Writer" of Efthymiou). 

Figure \ref{res:p2vec2} also shows that 
Limaye has higher Hits\# 
and lower Noise\# 
than Efthymiou.
This in some degree explains why P2Vec achieves more significant improvement on Limaye than on Efthymiou 
($0.081$ vs $0.05$ for the average accuracy gap of the $8$ classes in Figure \ref{res:p2vec2}; 
$25.3\%$ vs $4.3\%$ for the improvement by P2Vec in Figure \ref{res:p2vec1}).
Meanwhile, the low absolute value of Hits\# and Noise\# means P2Vec is quite sparse -- 
less than $0.3$ out of $422$ slots are none zero.
Sparsity reduces the training time and helps avoid over fitting.

\vspace{-0.15cm}
\vspace{-0.2cm}

\subsection{Ensemble}
As seen in Table \ref{res:ensemble}, both ensemble approaches are beneficial.
Ensemble I achieves higher accuracy than P2Vec and HNN on T2D-Te which comes from the same table set as the training data.
Ensemble II always achieves accuracy very close to the highest of P2Vec and HNN on all three testing sets, e.g., $0.650$ vs $0.655$ on Efthymiou. 
Ensemble I outperforms Ensemble II on T2D-Te, 
while Ensemble II outperforms Ensemble I on Limaye and Efthymiou.
Thus, we can apply Ensemble I in contexts where training and testing data come from the same source, 
and apply Ensemble II in contexts that need high robustness.
Considering Ensemble I re-trains a classifier with FC layer output of the trained HNN and P2Vec, and is more likely to be over-fitted to the training data, 
it is unsurprising to see its performance drop on Limaye and Efthymiou as the training data comes from T2Dv2.
This also indicates the difficulty of transferring learned table features and models between data sets.

\begin{table}[h!]
\scriptsize{
\centering
\begin{tabular}[t]{c|c|c|c}
\hline
Methods & T2D-Te & Limaye & Efthymiou    \\\hline 
 P2Vec &$0.939$ &$\bm{0.759}$ &$0.609$  \\
 HNN   &$\bm{0.962}$ &$0.728$ &$\bm{0.655}$  \\\hline
 Ensemble I (P2Vec + HNN)  &$\bm{0.966}$ &$0.697$ &$0.629$  \\
 Ensemble II (P2Vec + HNN) &$0.959$ &$\bm{0.746}$ &$\bm{0.650}$  \\\hline
\end{tabular}
\vspace{-0.2cm}
\caption{\footnotesize
Accuracy of P2Vec, HNN, and the ensemble approaches.
Both LR and MLP are used and the average is reported.
}\label{res:ensemble}
}
\end{table}
    
\subsection{Overall Result and Discussion}
As shown in Table \ref{res:baseline}, 
our method (HNN + P2Vec) dramatically outperforms Lookup-Vote and T2K Match that use lexical matching, and ColNet that uses deep learning,
when the training and testing data comes from the same table set (Local-70\%).
Its accuracy is $15.7\%$, $11.5\%$ and $5.0\%$ higher than Lookup-Vote,
$2.0\%$, $6.1\%$ and $6.4\%$ higher than ColNet,
on T2D, Limaye and Efthymiou respectively.
Although the assumption on training data would constrain applicability, 
the case that some columns have been annotated (e.g., by volunteers) while many more from the same source remain to be annotated is quite common.
On the other hand, when  trained on one table set (T2D-Tr) and transferred to another (Limaye and Efthymiou), the performance of HNN + P2Vec decreases but is still higher than ColNet. 
One cost sensitive solution for such a transfer setting is combining T2D-Tr with a small number of labeled columns from the testing set (Local-10\%);
its performance on Limaye is then $4.5\%$ and $12.3\%$ higher than Lookup-Vote and T2K Match respectively.

\vspace{-0.2cm}
\begin{table}[h!]
\scriptsize{
\centering
\begin{tabular}[t]{c|c|c|c}
\hline
Methods (Training Data) & T2D-Te & Limaye & Efthymiou    \\\hline 
 HNN + P2Vec (T2D-Tr) &\multirow{2}{*}{$\bm{0.966}$} &$0.746$ &$0.650$  \\
 HNN + P2vec (Local-70\%)   &  &$\bm{0.968}$& $\bm{0.865}$  \\
 HNN + P2vec (T2D-Tr + Local-10\% )    & - & $0.907$ & $0.697$  \\\hline
 Lookup-Vote  &$0.835$ &$0.868$ &$0.827$  \\
 T2K Match & 0$.772$ & $0.807$ & $0.612$   \\\hline
 ColNet (T2D-Tr) & \multirow{2}{*}{$0.947$} & $0.597$ & $0.619$   \\
 ColNet (Local-70\%) &  & $0.912$ & $0.813$   \\\hline
\end{tabular}
\vspace{-0.2cm}
\caption{\footnotesize
Accuracy of the baselines and our method under different training data settings. 
Local-$\lambda\%$ represents
randomly extracting $\lambda\%$ of a table set as training data, with the remainder as testing data. 
}\label{res:baseline}
}
\end{table}
\vspace{-0.1cm}

\noindent \textbf{Discussion.}
In the evaluation we first analyzed the impact of components of HNN.
Cell embedding by Att-BiRNN and column features by 
Conv filters over the target column 
achieve significant accuracy gains as expected, 
while row features by Conv filters over the row of the main cell 
have a positive impact on only one out of three testing sets,
which may be caused by varying table structures such as different column permutations.
%
Second, we evaluated P2Vec which is extracted by a KB lookup and query answering algorithm and includes information about potential relations between the target column and surrounding columns.
It achieves significant improvement, thus compensating for the above weak row features.
%
Third, we analyzed two ensemble approaches that combine P2Vec and HNN. 
They lead to better and more robust performance.
Finally we compared our method with some state-of-the-art baselines including those using deep learning (i.e., variants of HNN) and those using lexical matching (i.e., Lookup-Vote and T2K Match).
Our method 
significantly outperforms lexical matching
when the training data or a part of the training data comes from the same source as the testing data, but transferring the model trained on one table set to another totally different one for testing is still a big challenge.

\section{Related Work}\label{sec:related_work}
Most semantic table annotation works are based on lexical matching between table and KB \cite{venetis2011recovering,pham2016semantic,cafarella2018ten}.
State-of-the-art performance is achieved by jointly considering different matching tasks.
These methods include variants of probabilistic graphical models \cite{limaye2010annotating,mulwad2013semantic,bhagavatula2015tabel},
scoring models \cite{chu2015katara},
T2K Match \cite{ritze2015matching}, Table Miner \cite{zhang2017effective}, etc.
Performance also depends on the quality of the lexical index.
For example, the lookup service powered by the index of DBpedia Spotlight \cite{mendes2011dbpedia} 
can achieve good performance in cell to entity matching and column type annotation (i.e., Lookup-Vote) \cite{chen2019colnet}.
However, most of the above methods rely on table meta data for high performance, 
while lexical matching in principle fail to capture the contextual meaning of cell phrases.

Recently, with the development of deep learning, semantic embedding techniques like word2vec \cite{mikolov2013distributed} have been applied and 
methods that learn table features have been proposed.
Both \cite{efthymiou2017matching} 
and \cite{kunihiro2019meimei} utilize KB embedding.
The former explores the contextual semantics of an entity in the KB for disambiguation in cell to entity matching, 
while the latter accelerates searching and deals with the missing linkage in column to class matching with Markov Random Field.
\cite{luo2018cross}, \cite{nishida2017understanding} and \cite{chen2019colnet} all explore table feature learning with neural networks.
The former two learn cell features and locality features as our HNN, but deal with totally different problems.
\cite{luo2018cross} matches a cell to an entity in a different language, while \cite{nishida2017understanding} classifies the structure of a table.
In ColNet \cite{chen2019colnet we predict column types with a} different problem setting with unfixed candidate classes~and multiple binary classifiers. 
ColNet's architecture is~a special case of our HNN, namely word2vec + $\text{CNN}^{\text{c}}$ in Table~\ref{res:hnn}.
Briefly, learning the semantics of tabular data is promising, but still a big challenge \cite{thirumuruganathan2018data}.

\section{Conclusion and Outlook}\label{sec:conclution}
In this study we predict the semantic type of entity columns, 
using a hybrid neural network (HNN) for cell embedding and table feature learning,
and a property vector (P2Vec), extracted by KB lookup and query answering, for semantic features that represent potential inter-column relations.
We evaluated our method with DBpedia and three web table sets;
it is effective in most cases, and the overall performance exceeds the state-of-the-art in the supervised learning setting. 
We also considered generalisation across data sets, but this proved to be more challenging.
In the future we will apply our approach in an AI assistant for data analytics, and further investigate (permutation invariant) table feature learning.

\section{Acknowledgments}
We want to thank Chris Williams from University of Edinburgh for his constructive comments.
The work is supported by the AIDA project (UK Government's Defence \& Security Programme in support of the Alan Turing Institute), 
the SIRIUS Centre for Scalable Data Access (Research Council of Norway, project 237889),
the Royal Society,
EPSRC projects DBOnto, $\text{MaSI}^{\text{3}}$ and $\text{ED}^{\text{3}}$. 

\bibliographystyle{named}
\bibliography{reference}

\end{document}